\documentclass[preprint,aps,showpacs,superscriptaddress,nofootinbib,tightenlines]{revtex4}

\usepackage{amsfonts}
\usepackage{multirow}
\usepackage{mathrsfs}
\usepackage{graphicx}
\usepackage{amsmath}
\usepackage{amssymb}
\usepackage{bm}
\usepackage{bbm}
\usepackage{color}
\usepackage{hyperref}
\usepackage{slashed}
\usepackage{xparse}
\usepackage{upgreek}
\usepackage{comment}

\def\dr{\mathrm{d}}

\newcommand{\nn}{\nonumber}

\newcommand{\beq}{\begin{equation}}
\newcommand{\eeq}{\end{equation}}
\newcommand{\bqa}{\begin{eqnarray}}
\newcommand{\eqa}{\end{eqnarray}}
\newcommand{\bseq}{\begin{subequations}}
\newcommand{\eseq}{\end{subequations}}

\begin{document}

\title{\mbox{}\\[10pt]
Inclusive production of  fully-charmed ${\bm 1}^{\bm{+}\bm{-}}$ tetraquark at $\bm B$ factory}

\author{Yingsheng Huang\footnote{huangys@ihep.ac.cn}}
\affiliation{Institute of High Energy Physics, Chinese Academy of
  Sciences, Beijing 100049, China\vspace{0.2 cm}}
\affiliation{School of Physics, University of Chinese Academy of Sciences,
  Beijing 100049, China\vspace{0.2 cm}}

\author{Feng Feng\footnote{F.Feng@outlook.com}}
\affiliation{China University of Mining and Technology, Beijing 100083, China\vspace{0.2 cm}}
\affiliation{Institute of High Energy Physics, Chinese Academy of
  Sciences, Beijing 100049, China\vspace{0.2 cm}}

\author{Yu Jia\footnote{jiay@ihep.ac.cn}}
\affiliation{Institute of High Energy Physics, Chinese Academy of
Sciences, Beijing 100049, China\vspace{0.2 cm}}
\affiliation{School of Physics, University of Chinese Academy of Sciences,
Beijing 100049, China\vspace{0.2 cm}}

\author{\\Wen-Long Sang\footnote{wlsang@swu.edu.cn}}
\affiliation{School of Physical Science and Technology, Southwest University, Chongqing 400700, P.R. China}

\author{De-Shan Yang\footnote{yangds@ucas.ac.cn}}
\affiliation{School of Physics, University of Chinese Academy of Sciences,
Beijing 100049, China\vspace{0.2 cm}}
\affiliation{Institute of High Energy Physics, Chinese Academy of
Sciences, Beijing 100049, China\vspace{0.2 cm}}

\author{Jia-Yue Zhang\footnote{zhangjiayue@ihep.ac.cn}}
\affiliation{Institute of High Energy Physics, Chinese Academy of Sciences, Beijing 100049, China\vspace{0.2 cm}}
\affiliation{School of Physics, University of Chinese Academy of Sciences,
  Beijing 100049, China\vspace{0.2 cm}}

\date{\today}
\begin{abstract}
Inspired by the recent discovery of the $X(6900)$ meson at {\tt LHCb} experiment, we investigate the inclusive production rate of
the $C$-odd fully-charmed tetraquarks associated with light hadrons at the $B$ factory within the nonrelativistic QCD (NRQCD) factorization framework.
The short-distance coefficient is computed at lowest order in velocity and $\alpha_s$.
Employing the diquark-antidiquark model to roughly estimate the long-distance NRQCD matrix elements,
we predict the rate for inclusive production of the $1^{+-}$ $T_{4c}$ state and discuss the observation prospects
at {\tt Belle 2} experiment.
\end{abstract}

\maketitle


\section{Introduction}

Recently a narrow structure near $6.9\,\mathrm{GeV}$ in the di-$J/\psi$ invariant mass spectrum was reported by the {\tt LHCb} experiment,
with a global significance above $5\sigma$~\cite{Aaij:2020fnh}.
This somewhat unexpected discovery of the $X(6900)$ resonance has spurred a plethora of intensive theoretical investigations
to unravel its nature~(for an incomplete list of references, see \cite{liu:2020eha,Wang:2020ols,Jin:2020jfc,Yang:2020rih,Becchi:2020uvq,Lu:2020cns,Chen:2020xwe,
Karliner:2020dta,Zhao:2020nwy,Giron:2020wpx,Gordillo:2020sgc,Zhu:2020xni,Yang:2020wkh,Ke:2021iyh,Wan:2020fsk,Wang:2020wrp,Dong:2020nwy,
Gong:2020bmg,Albuquerque:2020hio,Albuquerque:2021erv,Zhu:2020snb,Dosch:2020hqm,Guo:2020pvt,Zhao:2020cfi}). The $X(6900)$ has been interpreted as $P$-wave fully-charmed tetraquark~\cite{liu:2020eha,Chen:2020xwe,Zhu:2020xni},
or the radially excited $S$-wave tetraquark~\cite{Lu:2020cns,Karliner:2020dta,Zhao:2020nwy,Zhao:2020cfi,Giron:2020wpx,Ke:2021iyh,Wang:2020ols,Yang:2020wkh,Zhu:2020xni}
or even the ground state $S$-wave tetraquark~\cite{Gordillo:2020sgc}. Alternatively, the $X(6900)$ is also suggested to be a
$\chi_{c0}\chi_{c0}$ or $P_c P_c$ molecular state~\cite{Albuquerque:2020hio,Albuquerque:2021erv},
$0^{++}$ hybrid~\cite{Wan:2020fsk}, the resonance formed in charmonium-charmonium scattering~\cite{Yang:2020rih,Jin:2020jfc},
or the kinematic cusp arising from final-state interaction~\cite{Wang:2020wrp,Dong:2020nwy,Gong:2020bmg,Guo:2020pvt}.
There has even been some attempts to tie $X(6900)$ with some beyond Standard Model scenario~\cite{Zhu:2020snb,Dosch:2020hqm}.

Unlike dozens of $XYZ$ states intertwined with the excited charmonia spectra
discovered during the past two decades, which necessarily contain light quark in their leading Fock component (for a recent review of $XYZ$, see Refs.~\cite{Guo:2017jvc,Liu:2019zoy,Ali:2017jda,Brambilla:2019esw}),
the $X(6900)$ is an entirely different exotic state, since its leading Fock component merely involves four heavy quarks.
Therefore, it is natural to envisage that, without pollution of the brown muck degrees of freedom,
the $X(6900)$ particle, among with other members in the $T_{4c}$ family,
should be much cleaner and amenable to study than its $XYZ$ cousins.
In particular, the asymptotic freedom of QCD may allow to address some dynamical features of the $T_{4c}$ family
within perturbative QCD thanks to $m_c\gg \Lambda_{\rm QCD}$.

Theoretical explorations of compact fully-heavy tetraquarks date back to 1970s~\cite{Iwasaki:1976cn,Chao:1980dv,Ader:1981db}.
Since then, the mass spectra and decay pattern of the fully-charmed tetraquarks (hereafter $T_{4c}$) have been extensively
investigated in various phenomenological models, such as quark potential models~\cite{Becchi:2020uvq,Lu:2020cns,liu:2020eha,Karliner:2020dta,Zhao:2020nwy,Zhao:2020cfi,Giron:2020wpx,Ke:2021iyh,Gordillo:2020sgc,Yang:2020rih,Jin:2020jfc}
and QCD sum rules~\cite{Chen:2020xwe,Wang:2020ols,Yang:2020wkh,Wan:2020fsk,Zhang:2020xtb}.
The studies on $T_{4c}$ production are relatively rare~\cite{Karliner:2016zzc,Berezhnoy:2011xy,Berezhnoy:2011xn,Becchi:2020mjz,Becchi:2020uvq,Maciula:2020wri,Carvalho:2015nqf,Gong:2020bmg,Goncalves:2021ytq},
most of which heavily rest upon some phenomenological ansatz such as quark hadron duality and color evaporation model.

It is intuitively appealing that, in order to produce the $T_{4c}$ state,
one has to first create four heavy quarks simultaneously at rather short spatial distance,
subsequently followed by nonperturbative hadronization process. The very first stage necessarily
involves hard momentum transfer, which can thus be accessed by perturbative QCD.
This is essentially the same physical consideration underlying the celebrated nonrelativistic QCD (NRQCD)
factorization approach to tackle ordinary quarkonium production.
Very recently, by drawing close analogy with quarkonium production, several groups have proposed to apply the NRQCD factorization approach
to study the $T_{4c}$ production at hadron colliders~\cite{Feng:2020riv,Ma:2020kwb,Zhu:2020xni} as well as $e^+e^-$ colliders~\cite{Feng:2020qee}.

To date, only the productions of the $0^{++}$ and $2^{++}$ $S$-wave $T_{4c}$ have been investigated in the aforementioned work,
mainly motivated by the $C$-even assignment of the $X(6900)$ by {\tt LHCb} experiment.
Nevertheless, there is a remaining $1^{+-}$ member in the $S$-wave $T_{4c}$ family, which has only received little attention thus far.
This $C$-odd tetraquark can decay into $J/\psi+\eta_c$ exclusively.
It is curious to speculate on where to look for this $C$-odd tetraquark.
In this work, our aim is to fill this gap by presenting a dedicated NRQCD analysis for the inclusive $1^{+-}$  $T_{4c}$ production at $B$ factory.
In particular, the production proceeds through $e^+e^-\to T_{4c}(1^{+-})+gg$, where charge conjugation invariance enforces that the fully-charmed
tetraquark to bear negative $C$ parity.
This study is especially of experimental interest, since analogous inclusive and exclusive quarkonium
production processes have already been extensively measured in {\tt Belle} experiments during the past two decades, exemplified by
$e^+e^-\to J/\psi+X$~\cite{Pakhlov:2009nj} and $e^+e^-\to J/\psi+\eta_c$~\cite{Abe:2002rb,Abe:2004ww,Aubert:2005tj}.
Moreover, since the $e^+e^-$ collision experiment has much cleaner background than {LHC},
the $B$ factory  might be an ideal place to look for the cousins of the $X(6900)$ particle.

The rest of the paper is organized as follows.
In section~\ref{2} we specify the NRQCD factorization formula for inclusive production of the $1^{+-}$ $T_{4c}$
associated with light hadrons.
In section~\ref{3}, we present the result for the short-distance coefficient (SDC) in factorization formula.
In section~\ref{4}, in the context of diquark-antidiqurak model, we give a rough estimate of
the value of the NRQCD long-distance matrix elements(LDMEs) based on quark potential model. We then make phenomenological analysis on the production
rate at $B$ factory and assess its observation prospect at {\tt Belle 2} experiment.
Finally in section~\ref{summary} we summarize.

\section{NRQCD factorization for $T_{4c}$ inclusive production\label{2}}

Our central goal is to predict the energy spectrum of the $1^{+-}$ tetraquark at $e^+e^-$ collider.
According to the spirit of the NRQCD factorization, we can express the differential cross section for $T_{4c}$ inclusive production as the sum
of the product of SDCs $\dr F_n$ and the LDMEs $\langle {\cal O}_n^{T_{4c}}\rangle$:
\beq
d\sigma(e^+e^-\to T_{4c}(E)+X)=\sum_{n} \dfrac{\dr F_{n}(E)}{m_c^8}(2M_{T_{4c}})\langle 0|\mathcal{O}_{n}^{T_{4c}}| 0\rangle,
\label{NRQCD:factorization:formula}
\eeq
where the sum is organized by velocity expansion.

In this work we concentrate on the $S$-wave $1^{+-}$ tetraquark. In the context of diquark picture, it is ready to
see the diquark and antidiquark pair should be in the $\mathbf{\bar 3}\otimes\mathbf{3}$ color state, consequently
Fermi statistics enforces the diquark/antidiquark to carry spin $1$. Bearing zero orbital angular momentum,
the diquark and anti-diquark then form total spin-$1$ tetraquark~\footnote{If they were in
$\mathbf{6}\otimes\mathbf{\bar 6}$ color state, the diquark/anti-diquark would be the spin-$0$ objects.
To form a spin-1 tetraquark, one must demand the orbital angular momentum between diquark and antidiquark to be $P$-wave,
hence suppressed by the velocity counting rule.}.
the lowest-order NRQCD production operator would not
involve any derivative. Within the diquark-antidiquark basis, the color-singlet production operator can be uniquely
defined as
\beq
\mathcal{O}_{\mathbf{\bar 3}\otimes\mathbf{3}}^{T_{4 c}} = \sum_{m_j,X} \mathcal{O}_{\mathbf{\bar 3}\otimes\mathbf{3}}^{i\dagger} |T_{4 c}(m_j)+X\rangle
\langle T_{4 c} (m_j) +X\vert \mathcal{O}^i_{\mathbf{\bar 3}\otimes\mathbf{3}},
\label{cs}
\eeq
where the magnetic quantum number represented by $m_j$, as well as the additional light hadronic states,
collectively denoted by $X$, are summed over.
Here the quadrilinear color-singlet NRQCD operator $\mathcal{O}_{\mathbf{3}\otimes\mathbf{\bar 3}}^{i}$ can be viewed
as the interpolating current bearing the same quantum number of the $1^{+-}$ tetraquark, whose explicit form
reads
\begin{align}
\mathcal{O}^{i}_{\mathbf{\bar 3}\otimes\mathbf{3}}= {i\over \sqrt{2}}\epsilon^{ijk}{\mathcal C}^{ab;cd}_{\mathbf{\bar 3}\otimes\mathbf{3}} \, \left(\psi_a^\dagger\sigma^j i \sigma^2\psi_b^{*}\right)\left(\chi_c^T i \sigma^2\sigma^k\chi_d\right).
\label{NRQCD:composite:operators}
\end{align}
Here $\psi$ and $\chi^\dagger$ are Pauli spinor fields that
annihilate the heavy quark and antiquark, respectively. $\sigma^i$ denotes Pauli matrix.
The Latin letters $i,j,k=1,2,3$ signify the Cartesian indices, whereas $a,b,c,d=1,2,3$ denote the color indices.
The color projection tensor in \eqref{NRQCD:composite:operators} is given by
\beq
\mathcal{C}^{ab;cd}_{\mathbf{\bar 3}\otimes\mathbf{3}}\equiv \left(\sqrt{1\over 2}\right)^2 \epsilon^{abe}\epsilon^{cdf}\frac{\delta^{ef}}{\sqrt{3}}=\frac{1}{2\sqrt{3}}(\delta^{ac}\delta^{bd}-\delta^{ad}\delta^{bc}).
\label{color:tensor}
\eeq
One can readily verify the NRQCD current in \eqref{NRQCD:composite:operators} has the prescribed
properties of the $1^{+-}$ state under $P$, $C$ transformations.

\section{Determining the short-distance coefficient\label{3}}

The SDCs in \eqref{NRQCD:factorization:formula} can be determined via the standard perturbative matching procedure.
Since these coefficients are insensitive to the long-distance nonperturbative dynamics, one is free to
replace the physical tetraquark state by a ``fictitious" tetraquark composed of
four free charm quarks, calculate both sides of \eqref{NRQCD:factorization:formula} using perturbative QCD and perturbative NRQCD,
then solve for SDCs.

We first use the standard trick to deduce the unpolarized production rate of $T_{4c}+gg$ in $e^+e^-$ annihilation
from the corresponding decay rate of a virtual photon:
\begin{align}
\mathrm{d}\sigma\left[e^+e^-\rightarrow T_{4c}(P)+g(k_1)g(k_2)\right]=&\dfrac{4 \pi\alpha}{s^{3/2}}{\mathrm{d}\Gamma\left[\gamma^*\rightarrow T_{4c}(P)+g(k_1)g(k_2)\right]},
\end{align}
where $\sqrt{s}$ denotes the center-of-mass energy of the $e^+e^-$ pair, and
$P,k_1,k_2$ denote the momenta of the tetraquark and two accompanying gluons.
For convenience, we introduce the following dimensionless ratios:
\beq
z={2 P^0 \over \sqrt{s}},\qquad x_1= {2 k_1^0\over \sqrt{s}},\qquad x_2= {2 k_2^0\over \sqrt{s}};\qquad r = {16m_c^2\over s}.
\eeq
The first three variables signify the energy fractions of the $T_{4c}$ together with two accompanying gluons, respectively,
which are subject to the constraint $x_1+x_2+z=2$ by energy conservation.

\begin{figure}
  \centering
  \includegraphics[width=0.5\textwidth]{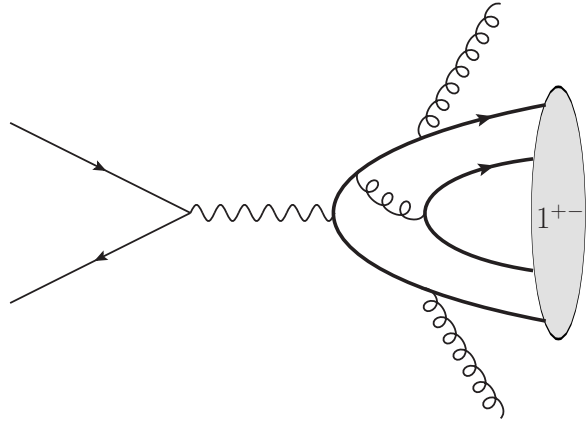}
  \caption{One of 392 Feynman diagrams for $e^+e^-\to T_{4c}(1^{+-})+gg$ at ${\cal O}(\alpha_s^4)$. }
  \label{feyndiag}
\end{figure}

At lowest order in $\alpha_s$, there are in total $392$ Feynman diagrams for $\gamma^* \to c c \bar{c}\bar{c} + gg$ in the perturbative QCD side,
one of which has been depicted in Fig.~\ref{feyndiag}. Among all the diagrams,
 $48$ diagrams in which two final-state gluons are emitted from a three-gluon vertex make vanishing contribution since
the $cc\bar{c}\bar{c}$ is in color octet.
Notice all topologies of diagrams start with ${\cal O}(\alpha_s^4)$, since $C$ conservation
demands that at least two gluons are emitted in the final state, and two charm quark lines must be connected through hard gluon exchange
to guarantee four charm quarks to move in the same direction, in order to have substantial probability to hadronize into a $T_{4c}$ state.

Since we are interested in the lowest order velocity expansion, we can simply assign each charm quark with momentum $P/4$, {\it i.e.}
equally partitioning the momentum of the fictitious tetraquark state. This is justified by the fact that the NRQCD current in
\eqref{NRQCD:composite:operators} contains no derivative.
To expedite the projection of the $cc\bar{c}\bar{c}$ state onto the fictitious tetraquark with prescribed color/spin/orbital quantum number,
we adopt a shortcut in the QCD-side calculation
by making the following substitution in the quark amplitude:
\begin{align}
  \bar u^a_i\bar u^b_j v^c_k v^d_l\to (\textsf{C}\Pi_\mu)^{ij}(\Pi_\nu \textsf{C} )^{lk}\mathcal{C}^{ab;cd}_{\bar{\mathbf{3}}\otimes\mathbf{3}} J^{\mu\nu}_{1}(\varepsilon), \label{sub}
\end{align}
where $\textsf{C}=i\gamma^0\gamma^2$ is the charge conjugate matrix,
$\Pi_\mu$ is the standard spin-triplet projector of bi-fermions~\cite{Feng:2020riv},
and the role of the projection tensor $J_{1}^{\mu \nu}(\varepsilon)=-i \epsilon^{\mu \nu \rho \sigma}\varepsilon_{\rho} P_\sigma/{\sqrt{2 P^{2}}}$
is to combine two spin-$1$ diquark-antidiquark pair into a $S$-wave spin-$1$ fictitious tetraquark state with polarization vector $\varepsilon^\rho$.
We simply take $P^2=M^2_{T_{4c}}\approx 16m_c^2$.

For the NRQCD-side calculation, one can also prepare a fictitious tetraquark state by setting all four
charm quarks at rest. The involved NRQCD matrix elements can be readily computed at lowest order in perturbation theory:
\bseq
\bqa
&& \left\langle\mathcal{T}_{\mathbf{\bar 3}\otimes\mathbf{3}}(m_j) \left|\mathcal{O}^i_{\mathbf{\bar 3}\otimes\mathbf{3}}\right|0 \right\rangle
 = 4\varepsilon^{i*}(m_j),
\label{vac:to:tetra:matrx:element:pNRQCD}
\\
&&  \langle 0|\mathcal{O}_{\mathbf{\bar 3}\otimes\mathbf{3}}^{{\cal T}_{4c}}| 0\rangle \approx
\sum_{m_j} \langle 0|  \mathcal{O}_{\mathbf{\bar 3}\otimes\mathbf{3}}^{i\dagger} |{\cal T}_{4 c}(m_j)\rangle
\langle {\cal T}_{4 c} (m_j) \vert \mathcal{O}^i_{\mathbf{\bar 3}\otimes\mathbf{3}}\vert 0\rangle = 48,
\label{inclu:prod:tetra:matrx:element:pNRQCD}
\eqa
\eseq
where $\varepsilon_{i}^{(m_j)}$ denotes the polarization tensor of the $1^{+-}$ state
with magnetic number $m_j$.
In the second line, we infer the inclusive production NRQCD matrix element from the vacuum-to-``tetraquark" matrix element \eqref{vac:to:tetra:matrx:element:pNRQCD}
by invoking vacuum saturation approximation (VSA).

To deduce the SDC affiliated with the differential production rate of $T_{4c}$ in \eqref{NRQCD:factorization:formula},
we need further integrate over the phase space integration of the gluons recoiling against $T_{4c}$~\footnote{
Upon squaring the QCD amplitude and summing over polarizations,  we use two different ways to conducting polarization sum
for external gluons. First we apply the Feynman gauge summation and including ghost contribution, alternatively we also choose
the polarization sum formula that only involve transverse polarizations without including ghost.  Both approaches yield identical
results.}.
We find the following formula for three-body phase space integration useful,
\beq
\int \!\! d\Phi_3 = \dfrac{s}{2(4\pi)^{3}}\int_{2\sqrt{r}}^{1+r}\dr z\int_{x_1^-}^{x_1^+}\dr x_1,
\label{three:body:phase:integral}
\eeq
where the integration boundaries of $x_1$ are
\beq
x_1^{\pm}=\dfrac{1}{2}(2-z)\pm\dfrac{1}{2}\sqrt{z^2-4r}.
\eeq

After some straightforward algebra, we obtain the intended SDC $\dr F_{\mathbf{\bar 3}\otimes\mathbf{3}}$ for differential
$T_{4c}$ energy distribution in \eqref{NRQCD:factorization:formula}. Unfortunately, the full analytical expression is too
lengthy to be presented in the text. As a compromise, we choose to present its limiting value near the
upper endpoint:
\begin{align}
\label{energy:spectrum:asym}
 { \dr F_{\mathbf{\bar 3}\otimes\mathbf{3}}\over \dr z}\bigg |_{z\to 1+r}
& =\dfrac{2^2\pi^3\alpha^2\alpha_s^4}{ 3^8s^2 (3-r)^2 (2-r)^2 (3+r) (6+r)}
\\
&\times \Big(550800+482112 \ln 2-803628 r-183168 r \ln 2+275616 r\ln r
\nn\\
&+27 \left(17856-16992 r-844 r^2+4764 r^3-779 r^4-336 r^5+70 r^6+r^7\right) \ln(2-r)
\nn\\
&+16 \left(-30132+11448 r-3897 r^2+8403 r^3-2489 r^4-475 r^5+166 r^6\right) \ln(3-r)
\nn\\
&+235854 r^2 +62352 r^2 \ln 2+85140 r^2 \ln r+62742 r^3-134448 r^3 \ln 2-263076 r^3\ln r
\nn\\
&-50316 r^4 +39824 r^4 \ln 2+60857 r^4 \ln r+2706 r^5+7600 r^5 \ln 2+16672 r^5 \ln r
\nn\\
&+1842 r^6-2656 r^6 \ln 2-4546 r^6 \ln r-27 r^7 \ln r\Big).
\nn
\end{align}

We can also obtain the integrated production rate for $e^+e^-\to T_{4c}+X$ by integrating \eqref{NRQCD:factorization:formula}
over $z$. To obtain the closed form, we choose to interchange the order of integration over $x_1$ and $z$ in
\eqref{three:body:phase:integral}.
The resulting SDC for the integrated cross section is still too lengthy to be presented here.
However, it is enlightening to present a compact asymptotic expression in the high energy limit $\sqrt{s}\gg 4m_c$:
\begin{align}
F_{\mathbf{\bar 3}\otimes\mathbf{3}}\big |_{r\to 0} & = \dfrac{\pi^3\alpha^2\alpha_s^4 }{2^23^8s^2}\Bigg[48(288\ln 3-167)
\ln\left({s\over 16 m_c^2}\right) - 417996\mathrm{Li}_2\left(\dfrac{1}{3}\right)-3744\text{Li}_2\left(\frac{3}{8}\right)
\nn\\
&+43005 \pi^2-386712+98082\ln^2 3+34128\ln^2 2+486032\ln 2+55296\ln 2\ln 3
\nn\\
&-218456\ln 3+11232\ln 2\ln 5-3744\ln 3\ln 5+167920\coth^{-1}2\Bigg].
\label{int:X:section:asym}
\end{align}
It is interesting to observe that at very high energy, the cross section decreases as
$\ln s/s^2$ asymptotically.

\section{Phenomenology\label{4}}

In this section, we proceed to a assess the observation prospect of the $1^{+-}$ tetraquark at {\tt Belle 2} experiment.
With the explicit knowledge of the desired SDC at hand, we still need a key ingredient,  {\it e.g.},
the nonperturbative NRQCD matrix element, in order to make a concrete phenomenological prediction of $T_{4c}$ production at the $B$ factory.
The ideal tool to conduct a model-independent prediction for the LDME would be lattice NRQCD simulation,
which, unfortunately, is unavailable at present. Therefore we must appeal to phenomenological models to infer the value of LDME.
For simplicity, we just employ a naive diquark model to give an approximate estimation.
The nonperturbative vacuum-to-tetraquark NRQCD matrix element turns to be
\begin{align}
   \left\langle T_{\mathbf{\bar 3}\otimes\mathbf{3}} (m_j)\left|\mathcal{O}^i_{\mathbf{\bar 3}\otimes\mathbf{3}}\right|0\right\rangle
     \approx
  \frac{\varepsilon^{i*}(m_j)}{2\pi^{3/2}} R_{D}^{2}(0) R_{T}(0),
\end{align}
where $R_D(0)$ and $R_T(0)$ denote the
radial wave functions at the origin for the diquark/anti-diquark and the whole diquark-antidiquark cluster.

We then appeal to VSA to deduce the desired vacuum matrix element of NRQCD production operator as introduced in
\eqref{NRQCD:factorization:formula}:
\begin{align}
\sum_{m_j}\left|\left\langle T_{4c}(m_j) \left|\mathcal{O}_{{\mathbf{\bar 3\otimes\mathbf{3}},\,i}}\right|0\right\rangle\right|^2=\frac{3}{4\pi^3}\left|{R_D(0)}\right|^4\left|{R_T(0)}\right|^2.
\end{align}

In the phenomenological analysis, we take $\sqrt{s}=10.58\,\mathrm{GeV}$,
$m_c=1.5\,\rm GeV$, $\alpha\left(10.58\,\rm GeV\right)=1/130.9$~\cite{Bodwin:2007ga}, $\alpha_s(2m_c)=0.2355$~\cite{Chetyrkin:2000yt}.
For the nonperturbative input parameter, we
we choose the diquark wave function at the origin $R_{D}(0)=0.523\;\mathrm{GeV}^{3/2}$~\cite{Kiselev:2002iy}.
The radial wave function at the origin for the diquark-antidiquark system has been computed in potential models~\cite{Debastiani:2017msn,Berezhnoy:2011xy,Berezhnoy:2012bv}.
We adopt the value $R_T(0)=2.902 \;\mathrm{GeV}^{3/2}$ which results from the Cornell-type potential model~\cite{Debastiani:2017msn}~\footnote{
A caveat is that our estimation of the NRQCD LDME is subject to very strong model dependence. For example, the value of wave function at
the origin will vary significantly if one switches to the color Coulomb potential.
Moreover, if one attempts to estimate the LDME by solving the four-body Schr\"{o}dinger equation, one might obtain a value far greater than that predicted in diquark model.}.

\begin{figure}
    \centering
    \includegraphics{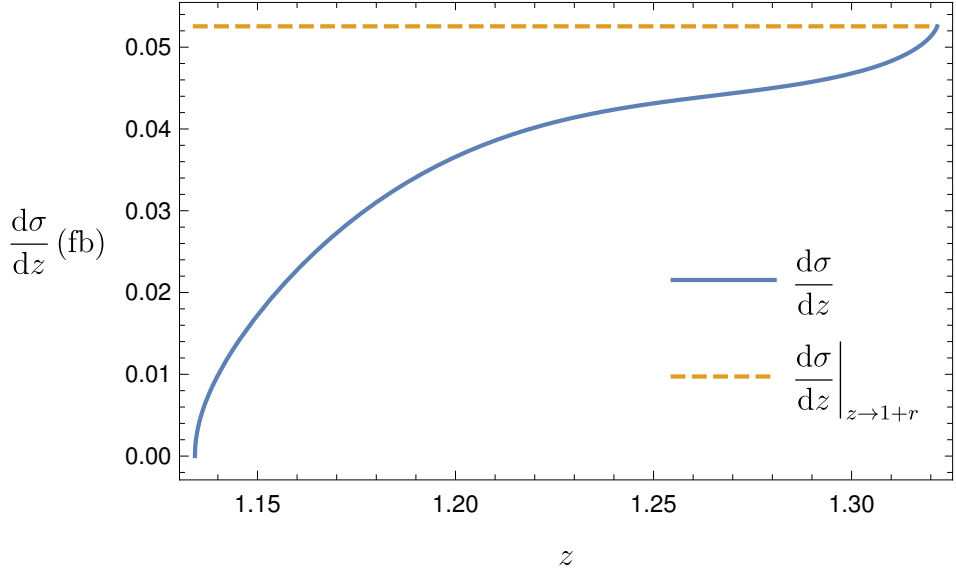}
    \caption{The energy distribution of $T_{4c}$ in the inclusive production from $e^+e^-$ annihilation at $\sqrt{s}=10.58\,\mathrm{GeV}$.
    The asymptotic value is taken from \eqref{energy:spectrum:asym}.
    }
    \label{Fig:energy distribution}
\end{figure}

\begin{figure}
    \centering
    \includegraphics{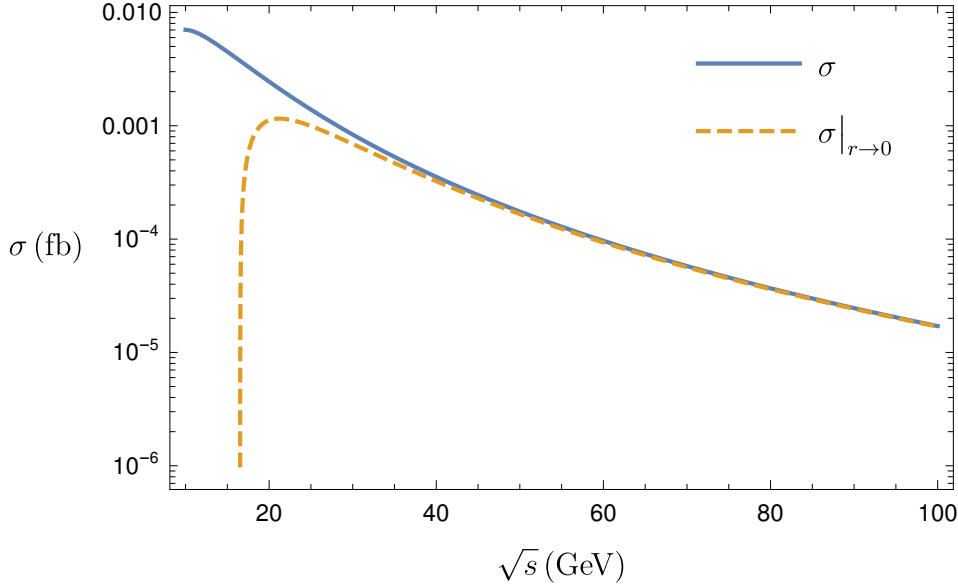}
    \caption{Integrated cross section as a function of center-of-mass energy.
    The asymptotic curve is taken from \eqref{int:X:section:asym}.
    }
\label{Fig:Int:X:section}
\end{figure}

In Fig.~\ref{Fig:energy distribution}, we plot the $1^{+-}$ tetraquark energy spectrum at $B$ factory energy.
We observe that the $1^{+-}$ tetraquark events favor to populate near the maximum allowed energy.
At $\sqrt{s}=10.58\,\mathrm{GeV}$, the integrated cross section is
\beq
\sigma(e^+e^-\to T_{4c}(1^{+-})+X) \approx0.0069\,\mathrm{fb}.
\label{num:prediction}
\eeq
This is an extreme tiny cross section.
Taking the projected integrated luminosity at {\tt Belle 2} to be $50\,\mathrm{ab}^{-1}$, we estimate there would be
$346$ events. Assuming one of the major decay channel to be $T_{4c}(1^{+-})\to J/\psi\eta_c$~\cite{Chen:2020xwe}, concerning the
small branching fraction of leptonic decay of $J/\psi$ and the experimental challenge to reconstructing $\eta_c$ unambigously,
as well as the copious background events,
the observation prospect of the $1^{+-}$ fully-charmed tetraquark appears to be rather pessimistic.

Nevertheless, the numerical prediction \eqref{num:prediction} may not need be taken too seriously, since
it is extremely sensitive to the nonperturbative input of the NRQCD LDME. Perhaps the more realistic estimation beyond
naive diquark model would give a much greater value, therefore the inclusive production rate may be enhanced by several orders of
magnitude. 
Perhaps a rigorous treatment based on four-body Schr\"{o}dinger equation will give more reliable estimate of the LDME.

In Fig.\ref{Fig:Int:X:section}, we also display the integrated cross section as function of $\sqrt{s}$.
One readily observes that, as long as $\sqrt{s}\ge 40 $ GeV, the asymptotic expression in 
\eqref{int:X:section:asym} starts to converge to the full result quite well.

\section{Summary}
\label{summary}

The recent discovery of the $X(6900)$ particle at {\tt LHCb} experiment has opened a new window toward studying
exotic hadrons, since it is likely the first genuine tetraquark composed of four charm quarks.
It might be naturally interpreted as a $0^{++}$ or $2^{++}$ $S$-wave tetraquark. In this work,
we study the inclusive production of the close cousin of the $X(6900)$,
a would-be fully-charmed $S$-wave tetraquark state with quantum number $1^{+-}$, in $e^+e^-$ annihilation.
In particular, we investigate the inclusive production rate of this $C$-odd $T_{4c}$ in association with light hadrons at {\tt Belle 2} experiment,
at the lowest order in NRQCD factorization approach.
We adopt a naive diquark-antidiquark cluster model to assess the encountered long-distance NRQCD matrix elements,
consequently predict a very tiny production rate, rendering its observation potential at {\tt Belle 2} experiment rather gloomy.
Nevertheless, we hope that a more realistic model may yield a much greater value of LDME so that the NRQCD prediction for the production rate
could be greatly enhanced. Needless to say, experimental search for fully-charmed tetraquarks at $B$ factory
will provide crucial guidance to our exploratory study.

\begin{acknowledgments}
 The work of Y.-S.~H., Y.~J. and J.-Y.~Z. is supported in part by the National Natural Science Foundation of China under Grants No.~11925506, 11875263,
 No.~ 12070131001 (CRC110 by DFG and NSFC).
The work of F.~F. is supported by the National Natural
Science Foundation of China under Grant No. 11875318,
No. 11505285, and by the Yue Qi Young Scholar Project
in CUMTB.
The work of W.-L. S. is
supported by the National
Natural Science Foundation of China under Grants No. 11975187 and the Natural Science Foundation of ChongQing
under Grant No. cstc2019jcyj-msxmX0479.
The work of D.-S.~Y. is supported in part by the National Natural Science Foundation of China under Grants No.~11635009.
\end{acknowledgments}




\clearpage


\begin{thebibliography}{100}

\vspace{3mm}

\bibitem{Aaij:2020fnh}
R.~Aaij \textit{et al.} [LHCb],
Sci. Bull. \textbf{65}, no.23, 1983-1993 (2020)
doi:10.1016/j.scib.2020.08.032
[arXiv:2006.16957 [hep-ex]].

%
\bibitem{liu:2020eha}
M.~S.~liu, F.~X.~Liu, X.~H.~Zhong and Q.~Zhao,
[arXiv:2006.11952 [hep-ph]].

%
\bibitem{Yang:2020rih}
G.~Yang, J.~Ping, L.~He and Q.~Wang,
[arXiv:2006.13756 [hep-ph]].

%
\bibitem{Wang:2020ols}
Z.~G.~Wang,
Chin. Phys. C \textbf{44}, no.11, 113106 (2020)
doi:10.1088/1674-1137/abb080
[arXiv:2006.13028 [hep-ph]].

%
\bibitem{Jin:2020jfc}
X.~Jin, Y.~Xue, H.~Huang and J.~Ping,
Eur. Phys. J. C \textbf{80}, no.11, 1083 (2020)
doi:10.1140/epjc/s10052-020-08650-z
[arXiv:2006.13745 [hep-ph]].

%
\bibitem{Lu:2020cns}
Q.~F.~L\"u, D.~Y.~Chen and Y.~B.~Dong,
Eur. Phys. J. C \textbf{80}, no.9, 871 (2020)
doi:10.1140/epjc/s10052-020-08454-1
[arXiv:2006.14445 [hep-ph]].

%
\bibitem{Becchi:2020uvq}
C.~Becchi, J.~Ferretti, A.~Giachino, L.~Maiani and E.~Santopinto,
Phys. Lett. B \textbf{811}, 135952 (2020)
doi:10.1016/j.physletb.2020.135952
[arXiv:2006.14388 [hep-ph]].

%
\bibitem{Chen:2020xwe}
H.~X.~Chen, W.~Chen, X.~Liu and S.~L.~Zhu,
Sci. Bull. \textbf{65}, 1994-2000 (2020)
doi:10.1016/j.scib.2020.08.038
[arXiv:2006.16027 [hep-ph]].

%
\bibitem{Albuquerque:2020hio}
R.~M.~Albuquerque, S.~Narison, A.~Rabemananjara, D.~Rabetiarivony and G.~Randriamanatrika,
Phys. Rev. D \textbf{102}, no.9, 094001 (2020)
doi:10.1103/PhysRevD.102.094001
[arXiv:2008.01569 [hep-ph]].

%
\bibitem{Giron:2020wpx}
J.~F.~Giron and R.~F.~Lebed,
Phys. Rev. D \textbf{102}, no.7, 074003 (2020)
doi:10.1103/PhysRevD.102.074003
[arXiv:2008.01631 [hep-ph]].

%
\bibitem{Wang:2020wrp}
J.~Z.~Wang, D.~Y.~Chen, X.~Liu and T.~Matsuki,
[arXiv:2008.07430 [hep-ph]].

%
\bibitem{Karliner:2020dta}
M.~Karliner and J.~L.~Rosner,
Phys. Rev. D \textbf{102}, no.11, 114039 (2020)
doi:10.1103/PhysRevD.102.114039
[arXiv:2009.04429 [hep-ph]].

%
\bibitem{Dong:2020nwy}
X.~K.~Dong, V.~Baru, F.~K.~Guo, C.~Hanhart and A.~Nefediev,
Phys. Rev. Lett. \textbf{126}, no.13, 132001 (2021)
doi:10.1103/PhysRevLett.126.132001
[arXiv:2009.07795 [hep-ph]].

%
\bibitem{Zhao:2020nwy}
J.~Zhao, S.~Shi and P.~Zhuang,
Phys. Rev. D \textbf{102}, no.11, 114001 (2020)
doi:10.1103/PhysRevD.102.114001
[arXiv:2009.10319 [hep-ph]].

%
\bibitem{Gordillo:2020sgc}
M.~C.~Gordillo, F.~De Soto and J.~Segovia,
Phys. Rev. D \textbf{102}, no.11, 114007 (2020)
doi:10.1103/PhysRevD.102.114007
[arXiv:2009.11889 [hep-ph]].

%
\bibitem{Zhu:2020xni}
R.~Zhu,
[arXiv:2010.09082 [hep-ph]].

%
\bibitem{Guo:2020pvt}
Z.~H.~Guo and J.~A.~Oller,
Phys. Rev. D \textbf{103}, no.3, 034024 (2021)
doi:10.1103/PhysRevD.103.034024
[arXiv:2011.00978 [hep-ph]].

%
\bibitem{Zhu:2020snb}
J.~W.~Zhu, X.~D.~Guo, R.~Y.~Zhang, W.~G.~Ma and X.~Q.~Li,
[arXiv:2011.07799 [hep-ph]].

%
\bibitem{Gong:2020bmg}
C.~Gong, M.~C.~Du, B.~Zhou, Q.~Zhao and X.~H.~Zhong,
[arXiv:2011.11374 [hep-ph]].

%
\bibitem{Wan:2020fsk}
B.~D.~Wan and C.~F.~Qiao,
[arXiv:2012.00454 [hep-ph]].

%
\bibitem{Dosch:2020hqm}
H.~G.~Dosch, S.~J.~Brodsky, G.~F.~de T\'eramond, M.~Nielsen and L.~Zou,
[arXiv:2012.02496 [hep-ph]].

%
\bibitem{Yang:2020wkh}
B.~C.~Yang, L.~Tang and C.~F.~Qiao,
[arXiv:2012.04463 [hep-ph]].

%
\bibitem{Zhao:2020cfi}
Z.~Zhao, K.~Xu, A.~Kaewsnod, X.~Liu, A.~Limphirat and Y.~Yan,
[arXiv:2012.15554 [hep-ph]].

%
\bibitem{Albuquerque:2021erv}
R.~M.~Albuquerque, S.~Narison, D.~Rabetiarivony and G.~Randriamanatrika,
[arXiv:2102.08776 [hep-ph]].

%
\bibitem{Ke:2021iyh}
H.~W.~Ke, X.~Han, X.~H.~Liu and Y.~L.~Shi,
[arXiv:2103.13140 [hep-ph]].

\bibitem{Guo:2017jvc} F.~K.~Guo, C.~Hanhart, U.~G.~Mei\ss{}ner, Q.~Wang, Q.~Zhao and B.~S.~Zou,
Rev. Mod. Phys. \textbf{90}, no.1, 015004 (2018)
[arXiv:1705.00141 [hep-ph]].

  \bibitem{Liu:2019zoy}
  Y.~R.~Liu, H.~X.~Chen, W.~Chen, X.~Liu and S.~L.~Zhu,
  Prog.\ Part.\ Nucl.\ Phys.\  {\bf 107}, 237 (2019)
  doi:10.1016/j.ppnp.2019.04.003
  [arXiv:1903.11976 [hep-ph]].

  \bibitem{Ali:2017jda}
  A.~Ali, J.~S.~Lange and S.~Stone,
  Prog.\ Part.\ Nucl.\ Phys.\  {\bf 97}, 123 (2017)
  doi:10.1016/j.ppnp.2017.08.003
  [arXiv:1706.00610 [hep-ph]].

  \bibitem{Brambilla:2019esw}
  N.~Brambilla, S.~Eidelman, C.~Hanhart, A.~Nefediev, C.~P.~Shen, C.~E.~Thomas, A.~Vairo and C.~Z.~Yuan,
  Phys. Rept. \textbf{873}, 1-154 (2020)
  [arXiv:1907.07583 [hep-ex]].


  \bibitem{Iwasaki:1976cn}
  Y.~Iwasaki,
  Phys.\ Rev.\ Lett.\  {\bf 36}, 1266 (1976).
  doi:10.1103/PhysRevLett.36.1266

  \bibitem{Chao:1980dv}
  K.~T.~Chao,
  Z.\ Phys.\ C {\bf 7}, 317 (1981).
  doi:10.1007/BF01431564

  \bibitem{Ader:1981db}
  J.~P.~Ader, J.~M.~Richard and P.~Taxil,
  Phys.\ Rev.\ D {\bf 25}, 2370 (1982).
  doi:10.1103/PhysRevD.25.2370

\bibitem{Zhang:2020xtb}
J.~R.~Zhang,
Phys. Rev. D \textbf{103}, no.1, 014018 (2021)
doi:10.1103/PhysRevD.103.014018
[arXiv:2010.07719 [hep-ph]].

  \bibitem{Berezhnoy:2011xn}
  A.~V.~Berezhnoy, A.~V.~Luchinsky and A.~A.~Novoselov,
  Phys.\ Rev.\ D {\bf 86}, 034004 (2012)
  doi:10.1103/PhysRevD.86.034004
  [arXiv:1111.1867 [hep-ph]].

\bibitem{Berezhnoy:2011xy}
  A.~V.~Berezhnoy, A.~K.~Likhoded, A.~V.~Luchinsky and A.~A.~Novoselov,
  Phys.\ Rev.\ D {\bf 84}, 094023 (2011)
  doi:10.1103/PhysRevD.84.094023
  [arXiv:1101.5881 [hep-ph]].

  \bibitem{Karliner:2016zzc}
  M.~Karliner, S.~Nussinov and J.~L.~Rosner,
  Phys.\ Rev.\ D {\bf 95}, no. 3, 034011 (2017)
  doi:10.1103/PhysRevD.95.034011
  [arXiv:1611.00348 [hep-ph]].

  \bibitem{Becchi:2020mjz}
  C.~Becchi, A.~Giachino, L.~Maiani and E.~Santopinto,
  Phys.\ Lett.\ B {\bf 806}, 135495 (2020)
  doi:10.1016/j.physletb.2020.135495
  [arXiv:2002.11077 [hep-ph]].

  \bibitem{Carvalho:2015nqf}
  F.~Carvalho, E.~R.~Cazaroto, V.~P.~Gonçalves and F.~S.~Navarra,
  Phys.\ Rev.\ D {\bf 93}, no. 3, 034004 (2016)
  [Phys.\ Rev.\ D {\bf 93}, 034004 (2016)]
  doi:10.1103/PhysRevD.93.034004
  [arXiv:1511.05209 [hep-ph]].

\bibitem{Maciula:2020wri}
R.~Maciu\l{}a, W.~Sch\"afer and A.~Szczurek,
Phys. Lett. B \textbf{812}, 136010 (2021)
doi:10.1016/j.physletb.2020.136010
[arXiv:2009.02100 [hep-ph]].

%
\bibitem{Goncalves:2021ytq}
V.~P.~Goncalves and B.~D.~Moreira,
Phys. Lett. B \textbf{816}, 136249 (2021)
doi:10.1016/j.physletb.2021.136249
[arXiv:2101.03798 [hep-ph]].


  \bibitem{Feng:2020riv}
  F.~Feng, Y.~Huang, Y.~Jia, W.~L.~Sang, X.~Xiong and J.~Y.~Zhang,
  [arXiv:2009.08450 [hep-ph]].

%
\bibitem{Ma:2020kwb}
Y.~Q.~Ma and H.~F.~Zhang,
[arXiv:2009.08376 [hep-ph]].


\bibitem{Feng:2020qee}
F.~Feng, Y.~Huang, Y.~Jia, W.~L.~Sang and J.~Y.~Zhang,
[arXiv:2011.03039 [hep-ph]].

\bibitem{Pakhlov:2009nj}
P.~Pakhlov \textit{et al.} [Belle],
Phys. Rev. D \textbf{79}, 071101 (2009)
doi:10.1103/PhysRevD.79.071101
[arXiv:0901.2775 [hep-ex]].

\bibitem{Abe:2002rb}
K.~Abe \textit{et al.} [Belle],
Phys. Rev. Lett. \textbf{89}, 142001 (2002)
doi:10.1103/PhysRevLett.89.142001
[arXiv:hep-ex/0205104 [hep-ex]].

\bibitem{Abe:2004ww}
K.~Abe \textit{et al.} [Belle],
Phys. Rev. D \textbf{70}, 071102 (2004)
doi:10.1103/PhysRevD.70.071102
[arXiv:hep-ex/0407009 [hep-ex]].

\bibitem{Aubert:2005tj}
B.~Aubert \textit{et al.} [BaBar],
Phys. Rev. D \textbf{72}, 031101 (2005)
doi:10.1103/PhysRevD.72.031101
[arXiv:hep-ex/0506062 [hep-ex]].

  \bibitem{Bodwin:2007ga}
  G.~T.~Bodwin, J.~Lee and C.~Yu,
  Phys. Rev. D \textbf{77}, 094018 (2008)
  doi:10.1103/PhysRevD.77.094018
  [arXiv:0710.0995 [hep-ph]].

\bibitem{Chetyrkin:2000yt}
K.~G.~Chetyrkin, J.~H.~Kuhn and M.~Steinhauser,
Comput. Phys. Commun. \textbf{133}, 43-65 (2000)
doi:10.1016/S0010-4655(00)00155-7
[arXiv:hep-ph/0004189 [hep-ph]].

   \bibitem{Kiselev:2002iy}
  V.~V.~Kiselev, A.~K.~Likhoded, O.~N.~Pakhomova and V.~A.~Saleev,
  Phys.\ Rev.\ D {\bf 66}, 034030 (2002)
  doi:10.1103/PhysRevD.66.034030
  [hep-ph/0206140].

   \bibitem{Berezhnoy:2012bv}
  A.~V.~Berezhnoy, A.~K.~Likhoded, A.~V.~Luchinsky and A.~A.~Novoselov,
  Phys.\ Atom.\ Nucl.\  {\bf 75}, 1006 (2012)
  [Yad.\ Fiz.\  {\bf 75}, 1067 (2012)].
  doi:10.1134/S1063778812040035

   \bibitem{Debastiani:2017msn}
  V.~R.~Debastiani and F.~S.~Navarra,
  Chin.\ Phys.\ C {\bf 43}, no. 1, 013105 (2019)
  doi:10.1088/1674-1137/43/1/013105
  [arXiv:1706.07553 [hep-ph]].


\end{thebibliography}
\end{document}